# Obesity Prediction with EHR Data: A deep learning approach with interpretable elements

Mehak Gupta[1,†], Thao-Ly T. Phan[3,4], H. Timothy Bunnell[1,2], Rahmatollah Beheshti[1,5,†]

[1] Computer and Information Sciences, University of Delaware, Newark, DE, USA
[2] Department of Biomedical Research, Nemours Alfred I. duPont Hospital for Children, Wilmington, DE, USA
[3] Department of Pediatrics, Nemours Alfred I. duPont Hospital for Children, Wilmington, DE, USA
[4] Department of Pediatrics, Thomas Jefferson University, Philadelphia, PA, USA
[5] Epidemiology Program, University of Delaware, Newark, DE, USA

† Email: mehakg@udel.edu (MG), rbi@udel.edu (RB)

**ABSTRACT**

Childhood obesity is a major public health challenge. Early prediction and identification of the children at a high risk of developing childhood obesity may help in engaging earlier and more effective interventions to prevent and manage obesity. Most existing predictive tools for childhood obesity primarily rely on traditional regression-type methods using only a few hand-picked features and without exploiting longitudinal patterns of children's data. Deep learning methods allow the use of high-dimensional longitudinal datasets. In this paper, we present a deep learning model designed for predicting future obesity patterns from generally available items on children's medical history. To do this, we use a large unaugmented electronic health records dataset from a large pediatric health system. We adopt a general LSTM network architecture which are known to better represent the longitudinal data. We train our proposed model on both dynamic and static EHR data. Our model is used to predict obesity for ages between 2-20 years. We compared the performance of our LSTM model with other machine learning methods that aggregate over sequential data and ignore temporality. To add interpretability, we have additionally included an attention layer to calculate the attention scores for the timestamps and rank features of each timestamp.

**CCS CONCEPTS**

• Computing methodologies • Machine learning • Machine learning approaches • Neural networks

**KEYWORDS**

• Childhood obesity • Electronic health records • Temporal data • Deep learning • Long short-term memory • Transfer learning

## 1 Introduction

Childhood obesity is a major public health problem across the globe as well as in the US. In 2019, the prevalence of obesity was 18.5% affecting almost 13.7 million US children and adolescents aged 18 or less [1]. Childhood obesity can continue into adulthood, which is known to be a major risk factor for chronic diseases such as diabetes, cancer, and cardiovascular diseases [2]. Preventing childhood obesity has been actively pursued in pediatric programs. However, decades of rigorous research have shown that prevention and management of obesity is not easy [3]. This is partly due to our limited understating of obesity and the complex interactions among a myriad of various factors, including biological and environmental ones, that are known to contribute to obesity. The motivation for this work is to use routinely captured EHR data in pediatric facilities to predict obesity without using data sources that are generally unavailable in pediatric

EHR datasets including maternity, pre-maternity, or lifestyle data. While flawed and incomplete, such EHR data is what providers are working with, and having a model embedded in the EHR data already available at a pediatric facility can help providers identify risk-patients without collecting extra data. Many predictors that other models use, such as maternal and pre-gestational data in more carefully crafted data collection studies are not available to the providers at pediatric facilities. Young children are particularly underdiagnosed with only 31% of overweight preschoolers versus 76% of overweight adolescents so identified [4]. Pediatric providers only infrequently use recommended Centers for Disease Control and Prevention's age and gender-specific BMI charts [5], explicitly designed to help screen for unhealthy weight trajectories. An objective tool might help providers identify high-risk patients and also drive the sensitive conversation with parents who do not recognize their children as overweight and may even fail to recognize the health risks of increased weight [6, 7].

In this study, we present a set of predictive models of childhood obesity using a longitudinal dataset of children derived from the electronic health records (EHR) of a large pediatric healthcare system. EHR data consists of clinical data along with its related temporal information. EHR data consists of a large number of features and not all features are recorded for each visit. This makes EHR data sparse (especially, when used in a one-hot encoding format for machine learning tasks). Also, EHR data is irregularly sampled, as there are no regular time-intervals between the patient's visits. The number of unique condition diagnosis, medications, procedures, and lab results collected in EHR datasets is generally huge. This leads to a very large feature (input) space for a prediction model, despite each visit having only a very small subset of total unique conditions, medications, procedures, and measurements recorded. The models we present in this study try to address these issues effectively. The EHR data used in this work consists of the records of patients' diagnosed conditions, prescribed medications, performed procedures, and recorded laboratory results in any visit. Compared to existing obesity predictive models in this domain, our model uses a much larger dataset (44 million rows with 68,003 unique patients) for training and considers a larger set of confounders for predicting outcomes. Our model is based only on the standard EHR data already available in many hospitals.

Besides only using the existing data on standard pediatric EHRs, another key difference between our models to the comparable models in this domain is considering the temporal changes in the children's health patterns. The major limitation of existing obesity models is twofold. First, available obesity models focus on single (or only a few) future point prediction for overweight or obese prediction (such as Santorelli et al. [8] at age 2, Weng et al. [9] at age 3, Redsell et al. [10] at age 5, Levine et al. [11] at age 5 and stratified by sex, Robson et al. [12] at age 5, Hammond et. al. [13] at age 5, Druet et al. [14] at age 7 and 14, Steur et al. [15] at age 8, Manios et al. [16, 17] at age 9 and 13, Pei et al. [18] at age 10, and Graversen et al. [19] at adolescence). These single-point prediction models cannot be generalized to predict the future BMI trajectories starting from various points in early childhood and adolescence. Obesity is prevalent in all age groups in childhood and adolescence. This makes the application of these models limited, as they cannot assist in predicting obesity status in other ages. The second limitation about existing models relates to using aggregated patterns instead of longitudinal patterns for developing the models [20]. For obesity, this is a major limitation, since rigorous research has shown that longitudinal patterns of obesity-related measures (such as body-weight) have a strong correlation with the future obesity patterns [21]. Similarly, a large body of research has shown that childhood obesity patterns are sensitive to different patterns of weight gain such that a more acute and rapid weight gain predicts a different severity of obesity versus a more gradual weight gain [21]. Aggregating EHR datasets (e.g., by calculating the average values) loses valuable knowledge from this type of datasets having time-series nature. The presented models in this work follow a recurrent neural network (RNN) architecture with long short-term memory (LSTM) cells and can learn a patient's representation from the temporal data collected over various visits of the patient. Additionally, as one of the major drawbacks of deep learning models like RNNs is the lack of interpretability, we use embedding weights on the input layer and softmax activations on LSTM layers of our networks to calculate the importance of the features and attention weights for each input timestamp. The importance score for the features and attention weights for timestamps are then used for visualizing different rankings. Apart from the temporal components, EHR data also contains static elements that do not change with every visit, such as sex, race, ethnicity, and zip code for each patient. We used a separate feed-forward network for the static data and merged its output with main architecture.

Our models can predict BMI (body mass index) values, defined as height in kg over height squared in meter, in ages 3 to 20. According to the US Center for Disease Control (CDC) [22], children in the highest 5th BMI percentile among the children of the same age and sex are considered obese. Having the estimated BMI values, we specifically look at the problem of classifying children as obese and non-obese.

The main contributions of this paper include presenting a predictive model of childhood obesity that 1) uses unaugmented EHR data, 2) can learn complex sequential patterns, and 3) is trained and evaluated on a very large sample (n=68,003 patients). Our models can predict obesity status at three different future time points (next 1, 2, and 3 years). Furthermore, we propose a mechanism to add interpretability to this model at both feature- and timestamp-level for the predictive task, which provides insights into important clinical events at individual and population levels. We perform comparisons between several machine learning techniques that ignore temporality and our-RNN based models that capture temporality in the data and show that our models can achieve better results. While we focus on an obesity-related problem, our approach should be also usable for studying similar problems using EHR data.

## 2 Related Work

Predictive models have recently shown a lot of promise in estimating future health outcomes in various biomedical applications; a trend similar to many non-biomedical domains [23-26]. Clinical predictive models are becoming prevalent and largely adopted due to an increase in the variability of medical data [27]. Until recently most of the clinical predictive models were primarily developed based on regression and logistic regression or other types of statistical analysis [20]. Traditional methods (including the machine learning ones) are not very effective in capturing the non-linear and temporal relationships in the complex EHR data. Recently, deep learning techniques have shown a lot of success in clinical predictive modeling [28].

Many of deep learning studies in this domain use RNNs, which refer to a special set of deep neural architectures used on sequential datasets. Unlike basic feedforward networks, RNNs can learn the long-term dependencies in temporal data by sharing parameters through the deep computational graphs. However, capturing long-term dependencies using RNNs generally faces the vanishing gradient problem where gradient values becoming too small. Hochreiter et al. [29] introduced the LSTM gating mechanism where the gradient can flow for long durations. LSTM gates can learn to keep important information and discard irrelevant ones from the previous time steps. This way, they can pass on the important information in the network for long durations. For additional details about RNN and LSTM architectures, we refer the readers to other references [29, 30].

Many clinical predictive models have been developed using RNNs to predict health problems like heart failure [31-33], diabetes [34], high blood pressure [35], and hospital readmission [36]. To name a few examples, Choi, et al. [31] use GRU for the multi-label prediction that predicts the diagnosis and medication codes for future visits. Choi et. al. [33] also uses RNNs with GRU for heart failure prediction, and Pham et. al. [34] use RNNs with LSTM for predicting readmission for diabetic and mental health patients. Maragathem, et al. [37] used the LSTM model with SiLU and tanh activation functions for heart failure prediction. Xu et al. [38] proposed multi-task RNN for multiple major adverse cardiovascular events (MACE) risk prediction on EHR data. Mei et al. [39] used raw EHR data to construct a "Deep Diabetologist" model with RNN for sequential data modeling. However, and despite the urgent need, there is not a lot of work done in the field of obesity predictive modeling leveraging large-scale datasets and advanced machine learning techniques. Most of the existing work relies on traditional machine learning methods. Example studies include using logistic regression or linear regressions [9, 10, 15, 19] and the random forest [13] where they used selected perinatal factors like BMI, birth weight, maternal weight, sociodemographic characteristics to predict obesity at certain age points in childhood and adolescence. A recent study has reviewed the machine learning models to predict childhood and adolescent obesity and compares these studies by also including an earlier version of our current work [40].

One of the major drawbacks of deep learning models is the lack of interpretability. The lack of interpretability reduces the value of predictive models, especially in the medical domain. If medical practitioners cannot understand how an outcome is predicted by a model, relying on the model's outcomes will not be practical [41]. Many attempts have been made recently to make sense of the outcome of these models. For instance, Bahdanau et al. [42] proposed the attention mechanism, which is a method originating from machine translation applications. This attention mechanism can improve interpretability at the time-level, as it gives attention scores to the timestamps. However, for multivariate time-series, we also need to consider feature importance at each timestamp. Zhang et al. [43] used a hierarchical attention mechanism by using a convolutional operation. Choi et al. [44] develop an interpretable model with two levels of attention weights learned from two reverse-time GRU models. Jin et al. [45] used two separate RNN networks to compute attention weights for analyzing EHR data. In our work, we continue the use of attention mechanisms to improve the interpretability of the RNN based models for multivariate time-series to obtain the importance score for the timestamps and additionally obtain the importance score for each feature in the timestamps.

## 3 Data

### 3.1 Dataset description

The EHR data used in this work was extracted from the Nemours Children Health System, which is a large network of pediatric health in the US, primarily spanning the states of Delaware, Florida, New Jersey, and Pennsylvania. The dataset is a portion of the larger PEDSnet dataset, containing EHR data from 8 major US Children's Health Systems [46]. Our data related to an extract of over two million distinct patients from Nemours EHR system with patient records dating from 2002 to the present. Inclusion criteria for the patients in the cohort used in this study included having: (i) at least 5 years of medical history, (ii) no evidence of type 1 diabetes, (iii) no evidence of cancer, sickle cell disease, developmental delay, or other complex medical conditions. In addition to including patients with obesity in our cohort, an equal number of normal weight patients were also selected by random sampling from the normal weight population. The analysis dataset was further screened for inconsistencies particularly those related to birthdates and measurement dates, dropping records with missing or implausible dates. The dataset was anonymized. All of the dates were skewed randomly per patient by +/- 180 days. The data access and processing steps were approved by the Nemours institutional review board. Each record in our dataset relates to one visit and captures the visit start and end time and all the condition, procedure, medication, and measurement variables recorded for that visit. It also contains demographic data for each patient. The medical codes are standardized terminologies of SNOMED-CT, RxNorm, CPT, and LOINC [47] for both clinical and demographic facts. Some additional details about the data are listed in Table 1.

Table 1: Characteristics of the cohort used in this study

| Name | Value |
| --- | --- |
| Total number of patients | 68,003 |
| Total number of visits | 34,96,559 |
| Avg. number of visits per patient | 51 |
| Number of females | 31,003 (45%) |
| Number of males | 37,000 (54%) |
| Age | 0-20, Mean = 5 |
| Race and Ethnicity | |
| White or Caucasian | 33244 |
| Black or African American | 25329 |
| Non-Hispanic or Latino | 58894 |
| Others | 17834 |

### 3.2 Data Representation and Preprocessing

The EHR data extracted for this study consisted of 44 million records with from 68,003 patients and 3,496,559 distinct visits. Medical codes in the data consisted of 20,300 condition, 10,167 procedure, 6,163 medication, and 7,693 measurement (lab-results) variables. On average, 22 condition, 34 procedure, 15 medication, and 49 measurement variables were recorded per patient in the data. Since the average number of medical variables recorded per patient was far lower than the total number of unique variables in the data, the dataset had a very sparse representation (especially, once we represent it using one-hot encodings for the trainings on machine learning models). We experimented with PCA (principal component analysis) and skip-gram methods for feature selection and dimensionality reduction [32]. Neither of these methods led to satisfactory results due to the very large size of our feature space. Finally, to reduce dimensionality, we followed a similar approach used in some other studies [48], we removed rare variables that did not have enough information and were absent in 98% of the population in the data. This resulted in 1850 medical variables with 635 condition, 407 procedure, 380 medication, and 428 measurement variables. Additionally, few measurement variables had inconsistent or incorrect values. More than 50% of entries in such variables contained text like "sample not collected properly" and "sample destroyed". We removed 113 such measurement variables that had missing and corrupt values. For any visit, if the same medical variable was recorded more than once, we took the latest value. There were several correlated features in the dataset like weight, height, BMI, obesity, BMI percentile. These variables had a different rate of missingness in the dataset. BMI, height, and weight were missing in 25%, 25%, and 7% of the visits in the final cohort respectively. Obesity was a condition variable that was only recorded if a patient was screened for obesity in that visit. Considering the missing rates, and to ensure our dataset can capture as much as obesity-related information, we did not do any further feature selection for these correlated variables. We also removed 26 patients from the data, which did not have any BMI recorded throughout their medical history. The final dataset consisted of 68,003 patients, 33,34,047 visits, and 1737 medical variables.

We preprocessed the original data to represent all the condition and procedure variables as binary variables (1 if present and 0 if not recorded for the visit). Few medication variables were present as continuous variables in original data, where the values contain information about the amount of medication prescribed to a patient during a visit. However, many medication variables were present as binary variables. Therefore, after the preprocessing, we represented all of the medication variables as binary variables. Measurement variables in the cohort were present as continuous variables. These continuous variables were normalized to the 0-1 range during preprocessing for model training. We have used zero imputation for missing entries in the EHR data. In medical data, where missingness carries meaning, imputation needs to include sufficient information about the reason for missingness to avoid bias [49]. However, most imputation techniques assume data to be missing at random [49]. Zero imputation sets the missing values to 0, such that it did not contribute to the patient's risk. It also has the potential to improve model performance in settings where missingness carries meaning [50]. Moreover, deep learning models like LSTMs has shown that they learn from missingness patterns represented by zero imputation [51]. With zero imputation it is also possible for the new patients to be assessed without the need for an end user to impute the values.

As the EHR data consists of patient records as a sequence of visits with each visit containing various medical codes, we use an incremental representation, where we first represent medical codes using code-level representation. Then, we use a code-level representation of all medical codes for each visit record and represent visits using visit-level representation. Finally, we use visit-level representation for each visit and represent the patients using a patient-level representation.

**Code-level Representation**

In code-level representation, medical codes consist of all the unique condition, medication, procedure, and measurement variables in the complete data. We denote condition codes with the vector $C : \{c_1, c_2, \ldots, c_{|C|}\}$ with a size of $|C|$, medication codes with the vector $D : \{d_1, d_2, \ldots, d_{|D|}\}$ with a size of $|D|$, procedure codes with the vector $P : \{p_1, p_2, \ldots, p_{|P|}\}$ with a size of $|P|$, and measurement codes $M : \{m_1, m_2, \ldots, m_{|M|}\}$ with a size of $|M|$.

**Visit-level Representation**

We denote a visit at time $t$ as $V_t$, which is the concatenation of the condition, medication, procedure, and measurement code vectors. The size of $V_t$ is $|V| = |C| + |D| + |P| + |M|$. We represent condition, medication and procedure codes for visit $V_t$ as binary vectors $C_t \in \{0,1\}^{|C|}$, $D_t \in \{0,1\}^{|D|}$, and $P_t \in \{0,1\}^{|P|}$ respectively, where "1" represents the presence of the corresponding code for the visit $V_t$. All the measurement variables are represented by the corresponding continuous values $M_t \in \mathbb{R}^{|M|}$ for the visit $V_t$.

Fig. 1 depicts the visit-level representation of our EHR data. EHR data for each patient is the sequence of visit-level vectors for that patient.

**Patient-level Representation**

Patient-level representation is the sequence of the visit vectors for a patient. We denote patients $S$ : $\{s_1, s_2, ..., s_{|N|}\}$ as $S$, where the $i$-th patient $s_i$ with $T$ visits is represented with the matrix $s_i \in \mathbb{R}^{|T|*|V|}$.

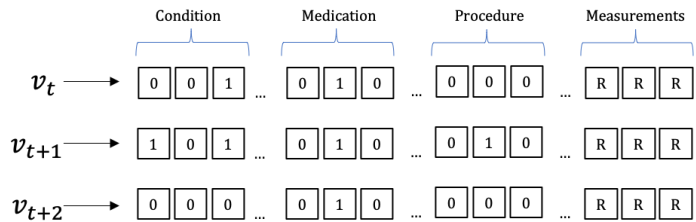

Fig. 1 – **Visit-level representation of the EHR data. Each visit $V_t$ is represented as a vector of condition, medication, procedure and measurement variables. The set of all visits for a patient forms a patient representation.**

In addition to the medical data, which changes with each visit, the EHR data also consists of static demographic data, which does not change with every visit. We represented the demographic variables, i.e., sex, race, ethnicity, insurance type, and zip code (indicating the approximate location of the patient) as categorical variables. Table 1 shows the distribution of race and ethnicity in the data.

As our cohort is extracted such that each patient has at least 5 years of data, sub-cohorts for every 5-year age range are created for the predictive models. Picking a longer range (than 5 years) would have resulted in a small number of patients as there were fewer patients who had records of more than 5 year (due to various reasons like having no visit or a provider change). We divided the complete cohort into 5-year ranges starting from each age from 0 to 15 years. This procedure resulted in 16 age ranges for the models (models for the ages of 0 to 5, 1 to 6, …, and 15 to 20). For every 5-year data, we then used a fixed observation window of the initial 2 years and predicted obesity for 1, 2, and 3 years in the future. This way, we ended up with 48 sub-cohorts by creating 3 sub-cohorts for each of the 16 5-year time-periods. For each of the 48 models, we used only those patients with at least one visit in the observation and one visit in the prediction window. Fig. 2 depicts how we created the sub-cohorts, as well as the observation and the prediction windows for each sub-cohort. The models are trained on data in the observation window to predict the future BMI values in the respective prediction window.

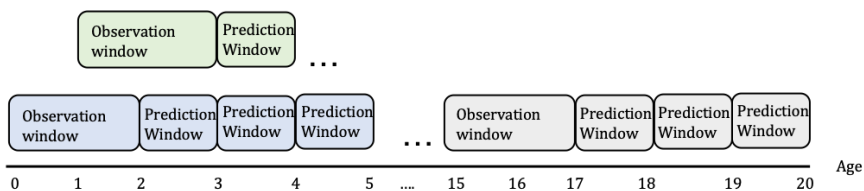

Fig. 2 – **Observation and prediction window design for each sub-cohort. For each observation window, there are 3 different prediction windows shown using same color.**

## 4 Method

### 4.1 LSTM model

After obtaining all the sub-cohorts as explained in Section 3.2, we transformed the data so that it can be given as input to the LSTM model. In general, visits have irregular time intervals and each patient has a different number of visits. To transform these irregularly spaced and unequal number of visits, we combined the visit data over a small fixed time window resulting in an equal number of time intervals. We combined visits over the 30-day time-periods of the 2-year observation window, resulting in (2*365)/30 ≈ 25 equally spaced sequences for each patient. Fig. 3 shows how new sequences are obtained from unequal and irregularly spaced input time sequences. Any condition, medication, and procedure variable observed at least once over the 30-day time-period is denoted by 1 in these new sequences. Continuous variables are averaged over the 30-day time-period. If there are no visits for a patient in any of the 30-day time-periods, the corresponding vector for that period contained all zeros. The zero vectors acted as padding to maintain an equal sequence length for the patients. In addition to the conditions, procedures, medications, and measurements, the time intervals between each visit sequence ($\Delta t$, capturing the time intervals between the non-empty sequences) were also added to the end of each visit vector. Adding time interval values has been shown to enrich the time-series input in similar studies [33]. Fig. S1 in the supplementary material shows the basic LSTM model architecture without components for interpretability.

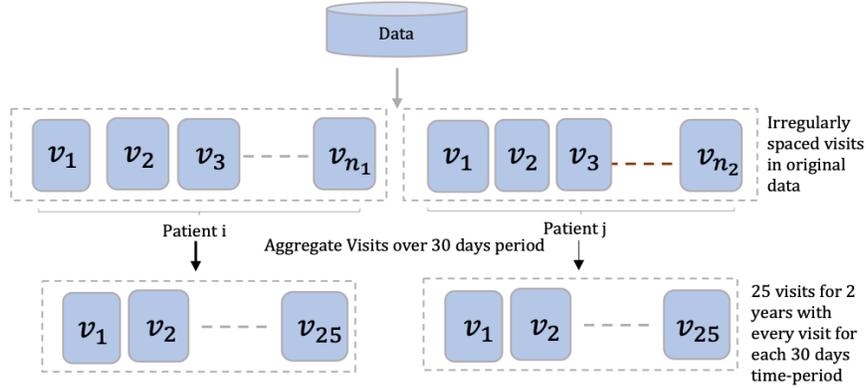

**Fig. 3 – Time sequences for the proposed LSTM model. Irregularly spaced visits for each patient (such as patient i and j) in a 2 -year period are mapped to 25 equally distributed intervals of 30 days length.**

We used LSTM layers to build our deep learning model. While deep learning models show superior performances compared to traditional machine learning models, they are difficult to interpret due to their so-called "black box" nature [52]. Such property may reduce the practicality of deploying these methods in medical domains. To mitigate such concerns, we enhanced the LSTM model by adding extra layers to improve its interpretability. Because of the mixed nature of our datasets, we have considered two levels of interpretability: time-level and feature-level interpretability. The time-level interpretability refers to ranking visits, and the feature-level interpretability refers to ranking the features present in the visits according to their importance in predicting the final output. Fig. 4 shows the complete architecture of the LSTM model with enhanced interpretability.

The architecture of our proposed model consists of various layers to transform the input data to predict future obesity status. $V = \{v_1, v_2, \ldots, v_t\}$ is a multivariate time-series input consisting of the visits for a patient observed over at timesteps $T = (1,2,\ldots,t)$. A visit $v_i$ is a $d$-dimensional vector consisting of all medical features for the visit $t$. The first layer in our architecture is an embedding layer, which is used to reduce the input dimension space from the $d$-dimensional to the $n$-dimensional space ($n<<d$). The output of the embedding layer is a multivariate time-series $X = \{x_1, x_2, \ldots, x_t\} \in \mathbb{R}^{t \times n}$. In our case study, we used $d = 1737$, and $n = 700$. This embedding layer is also used to add interpretability at the feature-level which we discuss in detail later. After the embedding layer, we used LSTM cells in two recurrent layers for training our model over the sequential visit-level (dynamic) data. The input to the LSTM layers is the multivariate time-series $X$. Each LSTM layer has a hidden layer dimension of size 700, which results in 3,922,800 trainable parameters. The output of the LSTM layers is another multivariate time-series in the $n$-dimensional space $H = \{h_1, h_2, \ldots, h_t\} \in \mathbb{R}^{t \times n}$. The softmax layer after the LSTM layers, generates attention scores $\{a_{11}, a_{12}, \ldots, a_{1n}; \ldots; a_{t1}, a_{t2}, \ldots, a_{tn}\} \in \mathbb{R}^{t \times n}$, where $a_{ij} \in (0,1)$. The average of all values in each vector $\{a_{i1}, a_{i2}, \ldots, a_{in}\}$ is then calculated to derive $\overline{a}_i$, where $i = \{1, 2, \ldots, t\}$. The $\overline{a}_i$ scores obtained from the softmax layer are used to add time-level interpretability which we discuss later. This softmax layer has 25 neurons, each related to one of the 25 timestamps, and adds additional 650 trainable parameters to our model. The weighted sum of the hidden states ($h_i$) and attention scores ($\overline{a}_i$) is used to obtain the final context vector $C$. There is also a separate feed-forward network for the demographic (static) data. The demographic input is $j$-dimensional vector $\{d_1, d_2, ..., d_j\} \in \mathbb{R}^j$, passes through two dense layers with 10 and 15 neurons and adds 50 and 165 trainable parameters respectively. The output of the LSTM layer is then concatenated with the output of this two-layer network. This concatenated output is then passed through one final dense layer consisting of 100 neurons with 71,600 trainable parameters for predicting the BMI values.

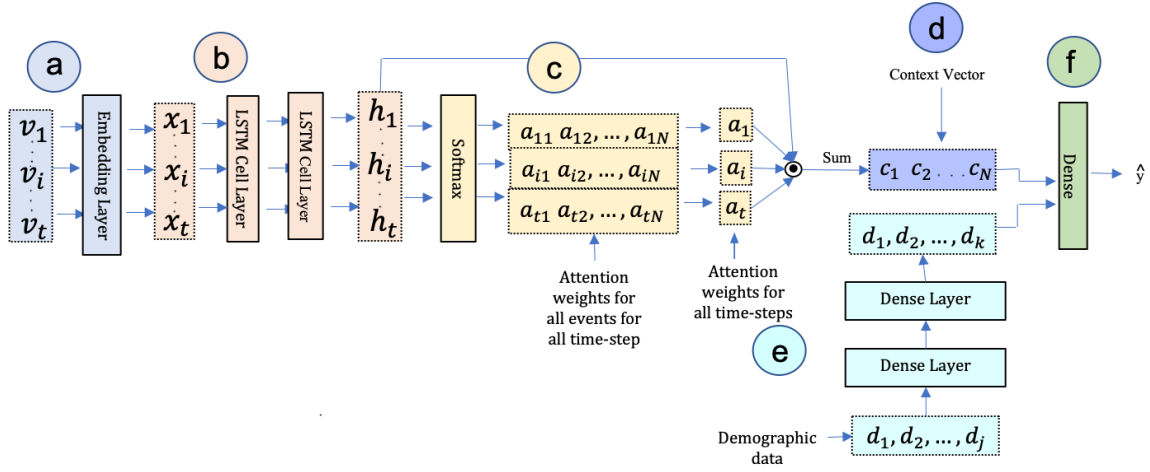

**Fig. 4 – Overview of proposed LSTM architecture including the elements to enhance its interpretability. (a) multivariate time-series input vector $V$ passes through the embedding layer. (b) The embedding layer's output $X$ passes through two LSTM layers. (c) A softmax layer generates attention weights. (d) A context vector is the weighted sum of the hidden state vectors $h_t$ and $a_t$. (e) A separate feed-forward network for the demographic data. (f) A fully connected layer to obtain the final output.**

**Time-level interpretability**

As mentioned previously, we added a softmax layer on top of the LSTM layers to compute the "attention score" for each timestamp. In general, each LSTM unit generates a hidden vector output $H = \{h_1, h_2, ..., h_i, ..., h_t\} \in \mathbb{R}^{t \times n}$, where $i = \{1, 2, ..., t\}$. The hidden state $h_i$ is computed by applying the non-linear transformation on input $x_i$ at time $i$ and the hidden state of previous time step $h_{i-1}$.

$$h_i \leftarrow LSTM(x_i, h_{i-1}) \qquad (1)$$

where $h_i$ is a $n$-dimensional vector, which is the same as the dimension of the hidden layer of LSTM network. We then calculated the attention score for each hidden state by using a softmax layer:

$$a_{i1}, a_{i2}, .....a_{ij}, a_{ij+1}, ....a_{in} = softmax(h_i) \qquad (2)$$

where $h_i$ is *n-dimensional* vector. $\{a_{i1}, a_{i2}, .....a_{ij}, a_{ij+1}, ....a_{in}\}$ is also an $n$-dimensional vector where $a_{ij} \in (0,1]$. To obtain a scalar attention score value for each hidden state, we took the average of all the values in the vector $\{a_{i1}, a_{i2}, ..... a_{ij}, a_{ij+1}, .... a_{in}\}$ as shown in Eq. 2.

$$\overline{a_i} = \left(\sum_{j=1}^{n} a_{ij}\right) \div n \qquad (3)$$

The value $\overline{a_i}$ is calculated for each hidden state $h_i$. These scores show importance scores for each $h_i$ and is used to visualize the visits that are given most importance by the LSTM layer. These scores are then used to compute the weighted sum of the hidden states (Eq. 4).

$$C = \sum_{i=1}^{T} \overline{a}_{i*} h_i \quad (4)$$

where $\overline{a_i}$ is a scalar value as shown in Eq. 3 and $h_i$ is an $n$-dimensional vector. $C$ is also an $n$-dimensional vector $\{c_1, c_2, \ldots, c_j, c_{j+1}, \ldots c_n\}$. The vector $C$ contains information of multivariate input time-series weighted by the importance score of each timestep [42].

**Feature-level interpretability**

As mentioned earlier, we added an embedding layer after the input layer to add feature-level interpretability. We employ simple method proposed in [53] to interpret medical code representations and enhance it using attention scores. We used the weights from the embedding layer to compute the importance score for the features in each timestep. The weight matrix of the embedding layer is of the form $W_i = w_1, w_2, \ldots, w_d \in \mathbb{R}^{n \times d}$ where $d$ is the dimension of the input time-series $v_i$ and $n$ is the reduced dimension of $x_i$ at timestep $i$. To obtain the feature ranking for each input feature in $v_i$ at the timestep $i$, we took the product of embedding of $v_i$ (weight matrix $W_i$ from the embedding layer) and the attention scores of $v_i$ obtained from the softmax layer, $b_i = \{a_{i1}, a_{i2}, \ldots a_{ij}, \ldots a_{in}\} \in (0,1]^n$. Eq. 5 shows the importance score calculation for features at timestep $i$:

$$s_i = W_i^T b_i \quad (5)$$

where $s_i$ is a $d$-dimensional vector containing the importance score for all input features.

## 4.2 Transfer Learning

We have additionally used transfer learning to enhance the model performance by learning from a larger dataset. While dividing our input data into the 48 sub-cohorts could improve its performance on learning specific age range patterns, this also meant reducing the input size for each of the models. This issue was especially more visible as the number of samples reduced gradually with the increasing age ranges. Reduction in the number of samples in pediatric datasets is common due to the higher rate of visits in earlier years of children's life. To improve the performance of the model, we used the complete dataset for all age ranges and created three initial models (instead of 48) for predicting obesity at 1 year, 2 years, and 3 years in the future respectively. After this, each of the 3 general models has been used as the basis for the 16 separate predictive models related to a similar prediction window.

# 5 Experiments

For training the models, we split data into 60:20:20 as training, validation, and test data. Data split is performed such that the proportion of obese and non-obese samples is the same in the training and test data as in original data. Table 2 shows the number of obese and non-obese samples in each sub-cohort.

We used two LSTM layers for all models and an Adadelta optimizer [54] with an initial learning rate of 0.05. We experimented with other learning rates (0.001, 0.01, 0.05, 0.1). We also tried different optimizer settings (RMSprop and Adam optimizer) with the learning rates mentioned before. We found the best train and validation error convergence with Adadelta and a learning rate of 0.05. Both L1 and L2 regularizations were used on the first LSTM layer. Models were trained on a Tesla V100-SXM2 32 GB GPU. We trained the models separately on different sub-cohorts using different observation and prediction age-windows as explained in Section 3.2. All models were trained on data in the observation window (first column in Table 2) to predict the future BMI value in the respective prediction window (second column in Table 2). The predicted BMI values were then used to classify each sample into obese and non-obese classes.

To evaluate the performance of our proposed model, we created two baseline models that follow the traditional methods that are commonly used in the literature. Such methods aggregate the patterns on the dataset while ignoring the temporality of the EHR data. We used linear regression and random forest as the baseline models for comparison. To do this, we aggregated data over all the visits in the observation window for each patient (corresponding to any of the input sub-cohorts). Target labels were the labels at the prediction age. Aggregation for binary medical codes for the conditions, medications, and procedures is performed such that each medical code represents the frequency of its occurrence over the 2 years, and for the continuous variables, we took the average over the 2 years. For the BMI and body weight, we took the maximum and latest recorded values in the observation window.

**Table 2 – Observation and prediction window setup for the 48 models, as well as the number of obese and non-obese samples in each model**

| Observation Window (Age year) | Prediction Window (Age year) | # of Obese Samples | # of Non-Obese Samples |
|---|---|---|---|
| 0-2 | 3, 4, 5 | 5556, 7492, 8192 | 27842, 25906, 25206 |
| 1-3 | 4, 5, 6 | 7382, 8081, 8307 | 25464, 24765, 24539 |
| 2-4 | 5, 6, 7 | 6697, 6878, 7014 | 19977, 19796, 19660 |
| 3-5 | 6, 7, 8 | 5697, 5788, 6216 | 16227, 16136, 15708 |
| 4-6 | 7, 8, 9 | 4840, 5194, 5679 | 13492, 13138, 12653 |
| 5-7 | 8, 9, 10 | 4428, 4813, 5117 | 11182, 10797, 10493 |
| 6-8 | 9, 10, 11 | 4085, 4368, 4574 | 9032, 8749, 8543 |
| 7-9 | 10, 11, 12 | 3773, 3941, 4047 | 7542, 7374, 7268 |
| 8-10 | 11, 12, 13 | 3273, 3354, 3387 | 6118, 6037, 6004 |
| 9-11 | 12, 13, 14 | 2671, 2729, 2692 | 4933, 4875, 4912 |
| 10-12 | 13, 14, 15 | 2116, 2078, 2078 | 3718, 3756, 3756 |
| 11-13 | 14, 15, 16 | 1507, 1502, 1530 | 2688, 2693, 2665 |
| 12-14 | 15, 16, 17 | 1052, 1059, 1087 | 1766, 1759, 1731 |
| 13-15 | 16, 17, 18 | 651, 665, 690 | 1044, 1030, 1005 |
| 14-16 | 17, 18, 19 | 250, 249, 260 | 358, 359, 348 |
| 15-17 | 18, 19, 20 | 55, 57, 55 | 87, 85, 87 |

## 5.1 Evaluation Results

We compared the performance of the baseline models and our proposed model. For the baseline models (linear regression and random forest), we used 10-fold cross-validation and report mean results over the complete data. For the LSTM models, we did not use cross-validations (due to heavy computing cost) and only report results on the test data. Fig. 5 shows the accuracy and AUC (area under the receiver operating characteristic curve) for all models separately based on the prediction window size. It compares the performance of 1) the proposed LSTM model with the interpretable elements trained using transfer learning, 2) random forest regressor, and 3) linear regression over different prediction windows separately. Refer to Table S1 in supplemental materials for an extended set of results of the proposed model. The additional results discussed in supplemental materials also include comparisons to another variation of the proposed model, which is the model without transfer learning. Fig. 6 demonstrates the effect of the size of the prediction window in predicting obesity at different prediction ages. It compares the sensitivity, PPV (positive predictive value), accuracy, and AUC of the proposed model.

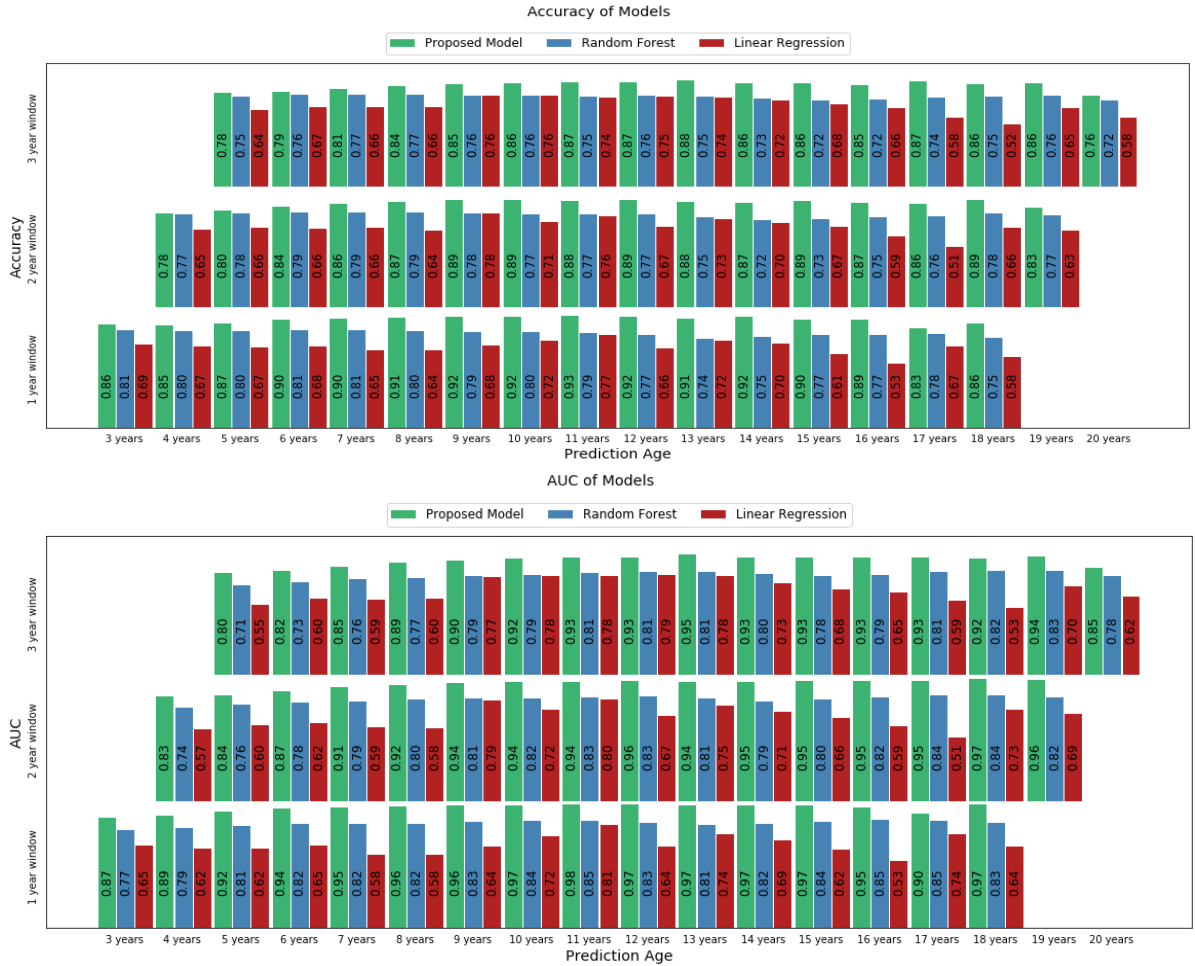

**Fig. 5 – Comparing accuracy and AUC results of 1) the proposed model, 2) random forest regressor, and 3) linear regression. Separate results are shown for different prediction windows at 1, 2, and 3 years in future.**

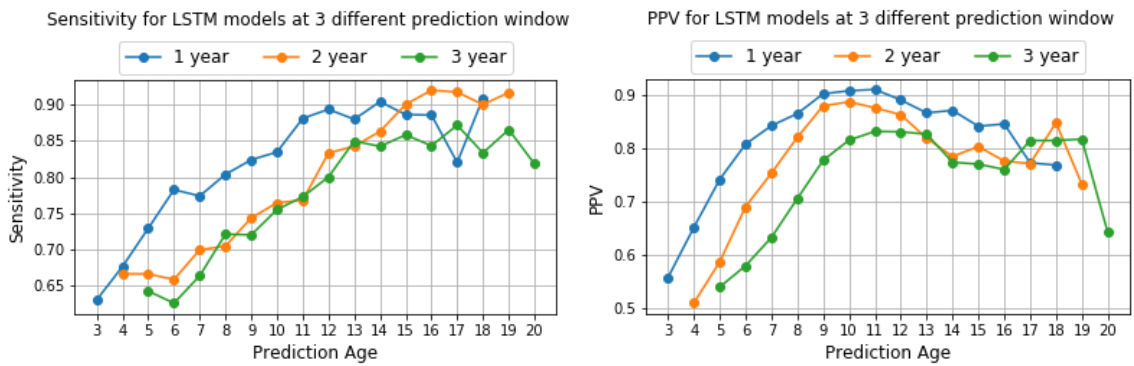

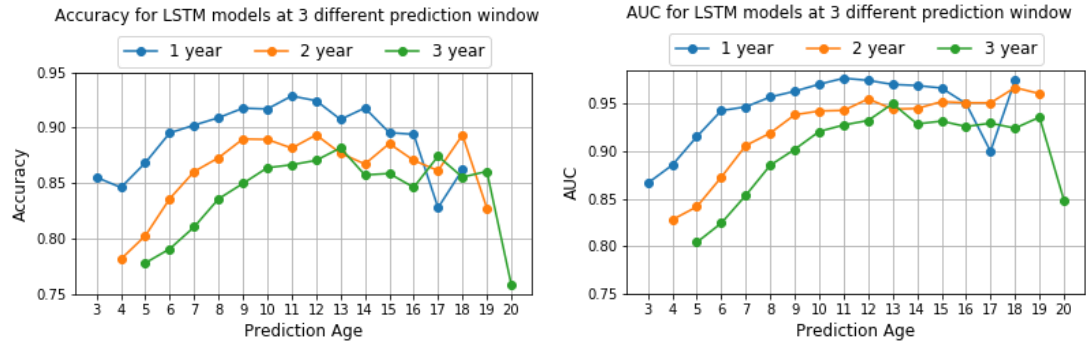

**Fig. 6 – Comparing prediction performances at different ages. Plots compare sensitivity, PPV, accuracy, and AUC of the proposed model across 1-, 2-, and 3-years in future, shown by blue, orange and green lines respectively.**

Additionally, we have computed the feature importance at both the individual and population levels. In Fig. 7, we ranked the features for the three most important visits for a sample patient to demonstrate the ability of our model to explain its predictions for any patient. To do this, we picked the top three visits with the highest attention weights and then ranked the features for those visits by calculating their importance score using Eq. 5. Refer to Fig. S3 in supplemental materials for the visualization of the attention weights given to input timestamps. We also ranked feature importance at the population level by averaging feature importance for the top 3 visits for all the individual samples. Table 3 shows the top 20 most important features at the population level.

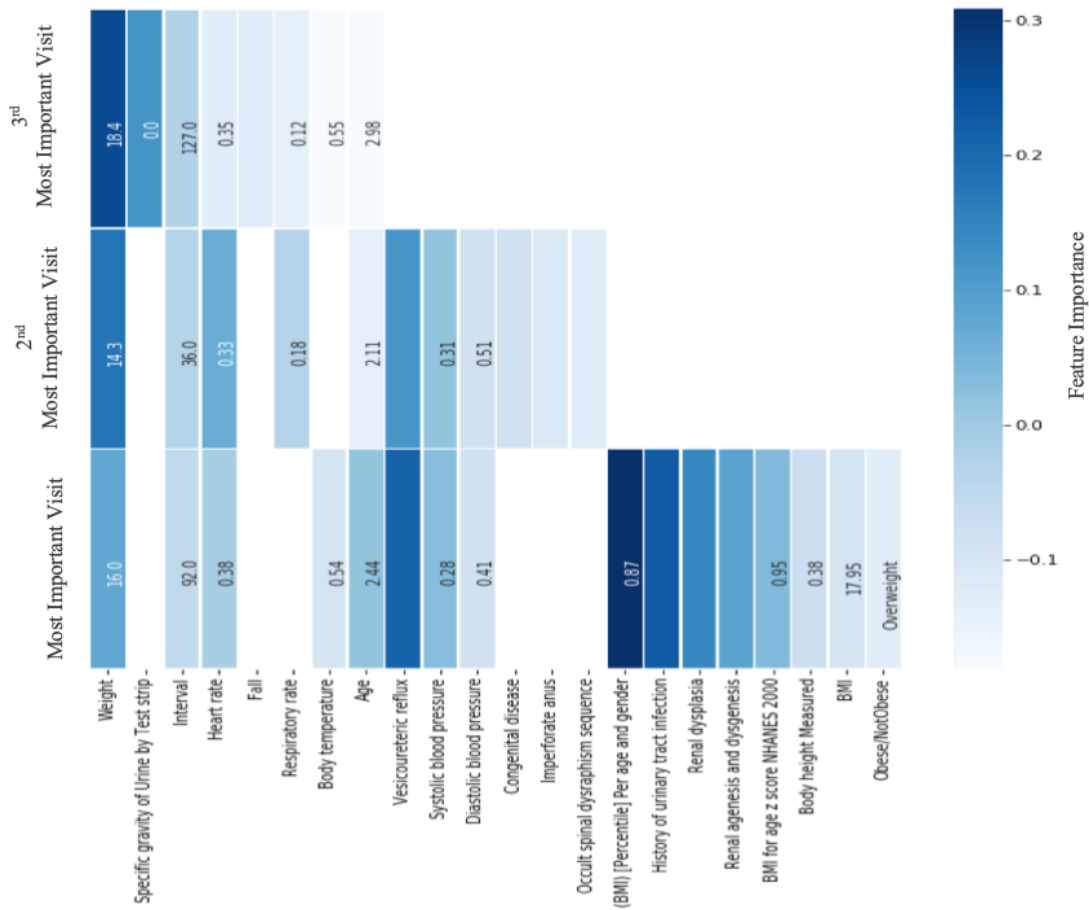

**Fig. 7 – Ranking of the features for the three most important visits (timestamps) for one patient (sample). Gradient bar on right shows the importance score for each feature (darker is more important). For the cells related to the measurements, each cell also contains the actual measurement's numerical value (such as weight and height). Non-measurement cells do not have a value.**

**Table 3. Feature rankings obtained by averaging the importance's score of the features from all the samples in the test data.**

| Feature Ranking | Feature Description | Feature Ranking | Feature Description |
|---|---|---|---|
| 1 | BMI [Percentile] Per age and gender | 11 | Abnormal weight gain |
| 2 | Obese/Non-obese Label | 12 | Anomaly of chromosome pair 21 |
| 3 | Allergic urticaria | 13 | Erythrocytes (#/volume) in body fluid |
| 4 | Childhood obesity | 14 | Obesity |
| 5 | Morbid obesity | 15 | Hyperactive behavior |
| 6 | Suspected clinical finding | 16 | Tachycardia |
| 7 | Achondroplasia | 17 | Requires respiratory syncytial virus Vaccination |
| 8 | MCH (Entitic mass) by Automated count | 18 | |
| 9 | Cholesterol in LDL/Cholesterol in HDL | 19 | $CO_2$|1712 |
| 10 | Hearing loss | 20 | pH of Blood |

We also compared the performance of our model with other existing childhood obesity prediction studies (Table 4). We compared the performance at the same prediction age with a 3-year window in our work as reported in existing works. If the existing work reported performance for a group of ages, then we took the average of performance at all the ages in the group at a 3-year window in our work.

**Table 4. Comparison of our work with existing work. Reg.: Regression**

| Existing Work | | | | Our Work |
|---|---|---|---|---|
| Author | Prediction Age | Model | Results | Results |
| Morandi et al. [55] | 7 and 16 years | Logistic Reg. | 0.78, 0.75 (AUC) | 0.85, 0.92 (AUC) |
| Druet et al. [14] | 7-14 years | Logistic Reg. | 0.77 (AUC) | 0.83 (AUC) |
| Weng et al. [9] | 3 years | Logistic Reg. | 0.75 (AUC) | 0.87 (AUC) |
| Redsell et al. [10] | 5 years | Logistic Reg. | 0.67 (AUC) | 0.80 (AUC) |
| Manios et al. [16] | 9-13 years | Logistic Reg. | 0.64 (AUC) | 0.92 (AUC) |
| Robson et al. [12] | 5 years | Logistic Reg. | 0.78 (AUC) | 0.80 (AUC) |
| Steur et al. [15] | 8 years | Logistic Reg. | 0.75 (AUC) | 0.88 (AUC) |
| Hammond et. al. [13] | 5 years | LASSO Reg. | 0.81 and 0.76 (AUC) for girls and boys | 0.80 (AUC gender combined) |
| Lingren et al. [56] | 1-6 years | Naïve Bayes | 0.90 (precision) and 0.44 (sensitivity) | 0.54 (precision) and 0.63 (sensitivity) |
| Zheng et. al. [57] | 14-18 years | Neural network | 0.84 (accuracy) | 0.85 (accuracy) |

## 5.2 Discussion

In this study, we have developed a new model for predicting childhood obesity patterns using EHR data. We employ an LSTM network and a separate feed-forward network to model dynamic (time-series) and static data in the EHR data. To transform the irregularly spaced and an unequal number of clinical visits, we combined the visits over a 30-day time window and obtained a regularly spaced and an equal number of clinical visits for each patient. We experimented with different window sizes of 6 months, 3 months, 30 days, and 15 days, and found out that a 30-day window size shows the best performance in capturing the variations in clinical trajectories of patients for predicting obesity. Notably, we achieved an AUC of 0.80, 0.93, and 0.92 using a 3-year input window for predicting obesity at 5 years (preschool), 11 years (prepubertal) and 18 years (post-pubertal).

As shown in Fig. 5, the performance of the proposed model was noticeably better than the two baseline models, i.e., linear regression and random forest. This shows that the RNN architecture improves the prediction performance by taking into consideration the temporality of the data, a property that traditional methods do not have. The existing body of research shows that data temporality is important to capture weight gain trajectories and other medical histories over time [21]. Moreover, transfer learning helps to further improve the performance of the models for the sub-cohorts with a low number of samples by learning from the samples in other sub-cohorts. For instance, prediction at age 20 using a 3-year window achieves accuracy and AUC of 0.45 and 0.55 respectively without using the transfer learning technique. However, when transfer learning is used, we achieved accuracy and AUC of 0.76 and 0.85 respectively. Refer to Fig. S2 for a comparison of the results obtained from training the proposed model with and without transfer learning. Here, transfer learning helps to improve the performance of the models on each sub-cohort by learning from the samples of other sub-cohorts.

As Fig. 6 shows, the closer the observation window is to the prediction time, the better the performance of the model is. This means that the prediction results obtained using a 1-year prediction window are better than the prediction results obtained on a 2-year window, which are in turn better than the prediction results obtained on a 3-year prediction window. For both accuracy and AUC, all of the plots show a semi-bell-shaped curve. In the beginning, the performance of the model increases, and then it starts to decrease after a certain age. One reason for observing such a pattern could be lower number of samples and the visits for each patient. As an example case, there is a sharp decrease in performance for prediction at the age 20. For the prediction at the age 20, our model uses an observation window of 15-17 years of age, with a very small number of samples. This small number of samples in the observation window of 15-17 years has more impact on the 3-year prediction window as compared to the 1-year and 2-year prediction window.

Besides superior predictive performance, we have also included interpretability capabilities in our proposed model. We have ranked the features in each visit to provide insights into the prediction results. We calculated the feature ranking over the complete dataset (with the samples that are predicted obese) to obtain the population-level feature ranking. As shown in Table 3, the most important population-level features in predicting childhood obesity as determined by our model were BMI, previous obesity levels (recorded as obese/non-obese), childhood obesity, morbid obesity, and obesity had the highest impact, matching to what reported in similar studies [58-60]. Other predictors of obesity with a lower prevalence that seemed to fit the existing work were achondroplasia [61, 62] and the anomaly of the chromosome of chromosome 21 [63]. Hyperactive behavior has a known association with obesity though the direction of causality is undetermined. Obesity is known to be among the main risk factors of high LDL (bad) cholesterol and low HDL (good) cholesterol [64]. MCH and Erythrocytes are also shown to be related to chronic inflammation of obesity [65], and $CO_2$ and pH levels of blood may be related to obesity hypoventilation syndrome [66, 67]. Another variable associated with obesity that is listed in Table 3 is the hearing loss, which has been show to be positively correlated with obesity [68].

Our results collectively show that feature ranking obtained from the proposed LSTM model can generate results that coincide with existing studies in the field of obesity research [69]. Similar to other predictive models, while our models cannot show any causation relationship, identifying the top features that the model has been using for its predictions may improve the explainability of our model. To further analyze the potential of the feature-level interpretability of our proposed model in understanding the roots of childhood obesity, we separately ranked the features for the patients who moved from non-obese to obese status, and the patients who were obese in the observation window and also stayed obese within the prediction window. Table S2 and Table S3 in supplemental materials show the list of top 20 features for both sub-groups. By comparing these top ranked features for the two sub-groups, we found that factors like prematurity (retinopathy of prematurity), acute tonsillitis and chronic adenoiditis were among the strong predictors in those who moved from the non-obese to obese status. Premature infants generally start with lower weight than their peers, but many gain weight rapidly [70]. Acute tonsillitis and chronic adenoiditis are strong predictors of obesity because of rapid weight gain after having tonsillectomy and adenoidectomy (which are the procedures to remove tonsils and adenoids) [71]. Factors like severe obesity are seen in patients who remain obese in both the observation and prediction windows. This is consistent with existing findings showing that those who develop obesity early on tend to have more a persistent obesity status than their counterparts who started with a normal weight [72]. Higher ranking of the LDL cholesterol feature in obese to obese patients may also show that those who develop obesity early also develop comorbidities of their obesity.

As shown in Table 4, our proposed deep learning model shows a better performance compared to traditional machine learning models. Also, unlike existing work that focuses on a single age-point to predict obesity, our model can predict obesity (BMI) patterns across a wide age-range, starting from early childhood to early adulthood. Ziauddeen et. al. [20] reported a list of existing studies on predicting childhood obesity, and all of these studies had used a logistic regression method on aggregated data. The data in these studies consisted of selected few variables such as maternal data and physical measurements of children. However, using an RNN-based architecture, our model was able to use temporal data and achieve a higher AUC compared to the similar prediction age points. Colmenarejo et. al. [40] have also reviewed existing work on obesity prediction that use logistic and other machine learning models. All the comparable studies are listed in Table 4. Some other studies related to obesity prediction have presented linear equations to estimate fat-free mass to access body fatness in children [73, 74]. These studies are not directly comparable to our work, as they only estimate the fat-free mass and not the total body weight. Additionally, most of such studies report single $R^2$ or RMSE scores over a wide range of ages such as 4-15 or 3-29, instead of at different ages (as done in our work). Another limitation of the linear models for obesity prediction is that they work on selected few predictor variables. Deep learning models as the one proposed in our work use a large number of predictor variables and can thus be extended to study the effect of other medical conditions in predicting obesity similar to the results shown in Table 3.

## 6 Limitations and Future Work

The current work is limited in several ways. Our model can predict a maximum of 3 years in the future. Long-term predictions require access to long-term medical histories. Since our data was collected in such a way that patients should have at least 5 years of data, we were only able to use a total 5-year window for performing supervised prediction tasks. In the future, we plan to explore other semi-supervised techniques like autoencoders to extend our prediction window with the complete data without dividing that into smaller datasets. The features in our EHR data contained only non-grouped features. Therefore, there was a large number of features and most of them were not recorded for a major portion of the population in the data. In the future, we aim to explore the opportunities to group medical variables and reduce the number of features in a more standardized manner. This may also help in obtaining better results from the feature-level interpretability process introduced here. Reducing the number of input features will also help to reduce the number of model parameters and will reduce its computational complexity. Currently, the results of feature-level interpretability consist of correlated features like existing obese/non-obese level, BMI, and weight. Our work can be extended by employing other feature selection techniques that can focus more on non-correlated features and possibly find high-risk features that are difficult for pediatricians to detect for predicting obesity. Other ways for extending our models include adding another attention layer after the initial dense layers for processing static (demographic) data and expanding our transfer learning process by using data from other healthcare systems.

## 7 Conclusion

In this study, we presented a new deep neural network architecture based on recurrent neural networks (RNNs) that combines both static and dynamic clinical features for predicting childhood obesity in the next one, two, and three years. Specifically, we have used RNNs with LSTM (long short-term memory) cells combined with a separate feedforward network for training our model. We trained the models using a large pediatric EHR (electronic health records) dataset. An additional transfer learning process was used to improve the performance of the models by using sub-cohorts of the complete dataset.

We achieved an AUC of 0.80, 0.93, and 0.92 for a 3-year window at 5 years (preschool), 11 years (prepubertal), and 18 years (post-pubertal). We also compared the recurrent model with other machine learning models for the task of predicting childhood obesity and found that our LSTM-based model demonstrates better performance compared to traditional machine learning models that ignore temporality of the data by aggregating the data. We also added interpretable elements to our proposed model, by ranking features in the top three most important visits during the two years of the training window for each patient. We have also calculated feature rankings for all samples in the data that transitioned from non-obese to obese status. Such a list of ranked features can provide further insights into the functionality of the models and demonstrate the features that contribute to childhood obesity most. Our models have the potential to be incorporated into existing EHR-based pediatric systems as a clinical decision support tool, helping both providers and families identify a child's risk for obesity earlier and make informed decisions about the need for earlier adoption of healthier lifestyles to prevent or treat childhood obesity.